\documentclass[prd,amsmath,aps,floats,amssymb, floatfix,
  superscriptaddress,nofootinbib,twocolumn,preprintnumbers]{revtex4-2}
\usepackage{tabularx}
\usepackage{amsmath,amssymb,natbib,latexsym,times}

\usepackage{booktabs}
\usepackage{multirow}

\usepackage{graphicx}
\usepackage{dcolumn}
\usepackage{bm}

\usepackage{url}
\usepackage{color}
\usepackage{comment}
\usepackage{mathrsfs}
\usepackage{hyperref}
\usepackage[dvipsnames]{xcolor}
\usepackage{orcidlink}
\usepackage{mathtools}

\newcommand{\mnras}{Monthly Notices of the Royal Astronomical Society}
\newcommand{\physrep}{Physics Reports}
\newcommand{\jcap}{Journal of Cosmology and Astroparticle Physics}

\renewcommand{\apj}{The Astrophysical Journal}

\newcommand{\aap}{Astronomy and Astrophysics}

\newcommand{\baas}{BAAS}

\begin{document}

\title[]{Testing small-scale modifications in the primordial power spectrum with Subaru HSC cosmic shear, primary CMB and CMB lensing 
}

\author{Ryo~Terasawa\orcidlink{0000-0002-1193-623X}}
\email{ryo.terasawa@ipmu.jp}
\affiliation{Kavli Institute for the Physics and Mathematics of the Universe (WPI), The University of Tokyo Institutes for Advanced Study (UTIAS), The University of Tokyo, Chiba 277-8583, Japan}
\affiliation{Department of Physics, The University of Tokyo, Bunkyo, Tokyo 113-0031, Japan}
\affiliation{%
Center for Data-Driven Discovery (CD3), Kavli IPMU (WPI), UTIAS, The University of Tokyo, Kashiwa, Chiba 277-8583, Japan}%

\author{Masahiro~Takada\orcidlink{0000-0002-5578-6472}}
\affiliation{Kavli Institute for the Physics and Mathematics of the Universe (WPI), The University of Tokyo Institutes for Advanced Study (UTIAS), The University of Tokyo, Chiba 277-8583, Japan}
\affiliation{%
Center for Data-Driven Discovery (CD3), Kavli IPMU (WPI), UTIAS, The University of Tokyo, Kashiwa, Chiba 277-8583, Japan}%

\author{Sunao~Sugiyama\orcidlink{0000-0003-1153-6735}}
\affiliation{Center for Particle Cosmology, Department of Physics and Astronomy, University of Pennsylvania, Philadelphia, PA 19104, USA}
\affiliation{Kavli Institute for the Physics and Mathematics of the Universe (WPI), The University of Tokyo Institutes for Advanced Study (UTIAS), The University of Tokyo, Chiba 277-8583, Japan}

\author{Toshiki~Kurita\orcidlink{0000-0002-1259-8914}}
\affiliation{Max Planck Institute f\"ur Astrophysik, Karl-Schwarzschild-Str. 1, 85748 Garching, Germany}
\affiliation{Kavli Institute for the Physics and Mathematics of the Universe (WPI), The University of Tokyo Institutes for Advanced Study (UTIAS), The University of Tokyo, Chiba 277-8583, Japan}

\begin{abstract}
Different cosmological probes, such as primary cosmic microwave background (CMB) anisotropies, CMB lensing, and cosmic shear, are sensitive to the primordial power spectrum (PPS) over different ranges of wavenumbers. In this paper, we combine the cosmic shear two-point correlation functions measured from the Subaru Hyper Suprime-Cam (HSC) Year~3 data with the {\it Planck} CMB data, and the ACT DR6 CMB lensing data to test modified shapes of the PPS at small scales, while fixing the background cosmology to the flat $\Lambda$CDM model. We consider various types of modifications to the PPS shape: the model with a running spectral index, the tanh-shaped model, the Starobinsky-type modification due to a sharp change in the inflaton potential, the broken power-law model, and the multiple broken power-law model.
Although the HSC cosmic shear data is sensitive to the PPS at small scales, we find that the combined data remains consistent with the standard power-law PPS, i.e., the single power-law model, for the flat 
$\Lambda$CDM background. In other words, we conclude that the $S_8$ tension cannot be easily resolved by 
modifying the PPS within the $\Lambda$CDM background.
\end{abstract} 

\maketitle

\section{Introduction}
The standard model of the universe, the flat $\Lambda$ Cold Dark Matter ($\Lambda$CDM) model has been successfully explains a variety of observations 
\citep[e.g.,][]{2013PhR...530...87W,2020moco.book.....D}. 
{The {\it minimal}} $\Lambda$CDM model {assumes adiabatic, Gaussian initial conditions and a single power-law form
for the power spectrum of primordial curvature perturbation.}
Although this simple model explains various observations fairly well, any features in the primordial power spectrum (PPS) would provide useful insights into the physics of inflation~\citep[e.g.][]{Planck2013_Inflation, Planck2015_Inflation,Chluba_2015, Planck2018_Inflation}.

For the shape of PPS, {\it Planck} CMB measurements have put tight constraints on the scale {range} $k \sim [10^{-4}, 0.2]~{\rm Mpc^{-1}}$ \citep[e.g.][]{Planck2013_Inflation, Planck2015_Inflation, Planck2018_Inflation}.
On smaller scale, constraint can be 
{derived from high-multipole CMB data}
(e.g., from ACT~\citep{2024JCAP...12..038H}), Ly-$\alpha$ forest~\citep{2025PhRvR...7a2018R}, $\mu$-distortions~\citep{2017PhRvD..95l1302N, 2019BAAS...51c.184C}, and future 21cm experiments~\citep{2025arXiv250102538N}.
In this paper, we focus on cosmic shear data, which refers to the coherent distortions of galaxy images 
{caused by weak gravitational lensing due to intervening large-scale structure} \citep[e.g.,][]{2017grle.book.....D}.
Since the large-scale PPS is well-constrained with the {\it Planck} data, 
we primarily focus on small-scale PPS features, which can be explored using cosmic shear data.
Cosmic shear is sensitive to the scale around $k \sim [10^{-2}, 1]~{\rm Mpc}^{-1}$~\citep[e.g.][]{Terasawa24}. Hence it bridges the gap between the scale probed by {\it Planck} CMB and $\mu$-distortions, which are directly sensitive to the scale around $k \sim 10^{3}~{\rm Mpc}^{-1}$~\citep{2012ApJ...758...76C,2013JCAP...02..036P,2013JCAP...06..026K,2014MNRAS.438.2065C}. 

Despite the success of the {minimal} $\Lambda$CDM model, recent {high-precision}
observations have indicated {potential}
discrepancies from {this}
model.
One such discrepancy is known as the $\sigma_8$ or $S_8$ tension,
{where $S_8$ characterizes the amplitude of matter clustering in the present-day universe.} 
This refers to the consistent lower values of $\sigma_8$ or $S_8$ in $\Lambda$CDM models inferred from large-scale structure (LSS) probes, 
compared to those inferred from the \textit{Planck}-2018 CMB measurements~\citep[see][for a recent review]{2022JHEAp..34...49A}.
Such large-scale structure probes that exhibit the $S_8$ (or $\sigma_8$)
tension include cosmic shear
\citep[e.g.,][]{HSC3_cosmicShearReal,HSC3_cosmicShearFourier,KiDS1000_CS_Asgari2020,DESY3_CS_Secco2022},
joint probe cosmology combining cosmic shear
and galaxy clustering
\citep[e.g.,][]{HSC3_3x2pt_ss,HSC3_3x2pt_ls,KiDS1000_3x2pt_Heymans2021,DESY3_3x2pt2022}
and redshift-space galaxy clustering \citep[e.g.,][]{2020JCAP...05..042I,2022JCAP...02..008C,2022PhRvD.105h3517K}. 
{Given the tension between the CMB and the cosmic shear data,}
{in this paper we examine whether}
a modification of PPS {can provide a better fit to both 
the {\it Planck} CMB and cosmic shear data
over the single power-law PPS model.}
A number of works tried to reconstruct the PPS or constrain the primordial features using CMB and/or LSS data~\citep[e.g.][]{2019PhRvR...1c3209B, 2021JCAP...10..081C, 2023JCAP...08..012M}.
{However,}
most of them fix the {background} $\Lambda$CDM model in the analysis. 
{{In contrast,}
we simultaneously vary the cosmological parameters along with the parameters characterizing the modified PPS model, {offering a more comprehensive exploration of possible solutions to the $S_8$ tension as well as testing the modification of the PPS.}
To perform our study, we use the cosmic shear data from the Subaru 
Hyper Suprime-Cam (HSC) Year~3 data, {\it Planck} CMB data, ACT DR6 CMB lensing data,
and the DESI Year~1 data of baryon acoustic oscillation (BAO) measurements.
However, note that we {assume} 
the background cosmology to be the flat $\Lambda$CDM model, while 
the background cosmological parameters are allowed to vary in the inference.}

Similarly to the $S_8$ tension, Ref~\cite{2025PhRvR...7a2018R} {showed}
that
combined \textit{Planck} CMB, BAO
and supernovae data, when  
analyzed under the $\Lambda$CDM model, are in $4.9\sigma$ tension with the eBOSS Ly-$\alpha$ forest in the inference of the linear matter power spectrum at $k \sim 1~h {\rm Mpc}^{-1}$ and redshift $z=3$.
They found running in the tilt of the PPS ($\alpha_s \sim -0.01$), as well as other model extensions (ultra-light axion dark matter or warm dark matter), can alleviate the tension. 
Since the cosmic shear probes similarly small scales ($k \sim [10^{-2}, 1]~{\rm Mpc}^{-1}$) but lower redshifts ($z \lesssim 1$), 
it
would serve as a complementary probe to test 
{if the negative running index}
preferred in the Ly-$\alpha$ analysis is {also favored by the cosmic shear data.}

We organize this paper as follows. In Section~\ref{sec:pps-theory}, we introduce the model for the PPS and discuss the effects on the cosmic shear signal.
In Section~\ref{sec:analysis}, we explain the details of the joint analysis of the cosmological datasets presented in this paper, and we will show the results in Section~\ref{sec:results}.
Section~\ref{sec:conclusion} is devoted to conclusion.

\section{Modification to the primordial power spectrum}
\label{sec:pps-theory}

In Section~\ref{subsec:templates},
we introduce models for the primordial power spectrum 
(PPS) considered in this paper.
We discuss the effects on the cosmic shear signal in Section~\ref{subsec:cosmic shear}.

\subsection{Template models of PPS}
\label{subsec:templates}

\begin{figure}
    \includegraphics[width=\columnwidth]{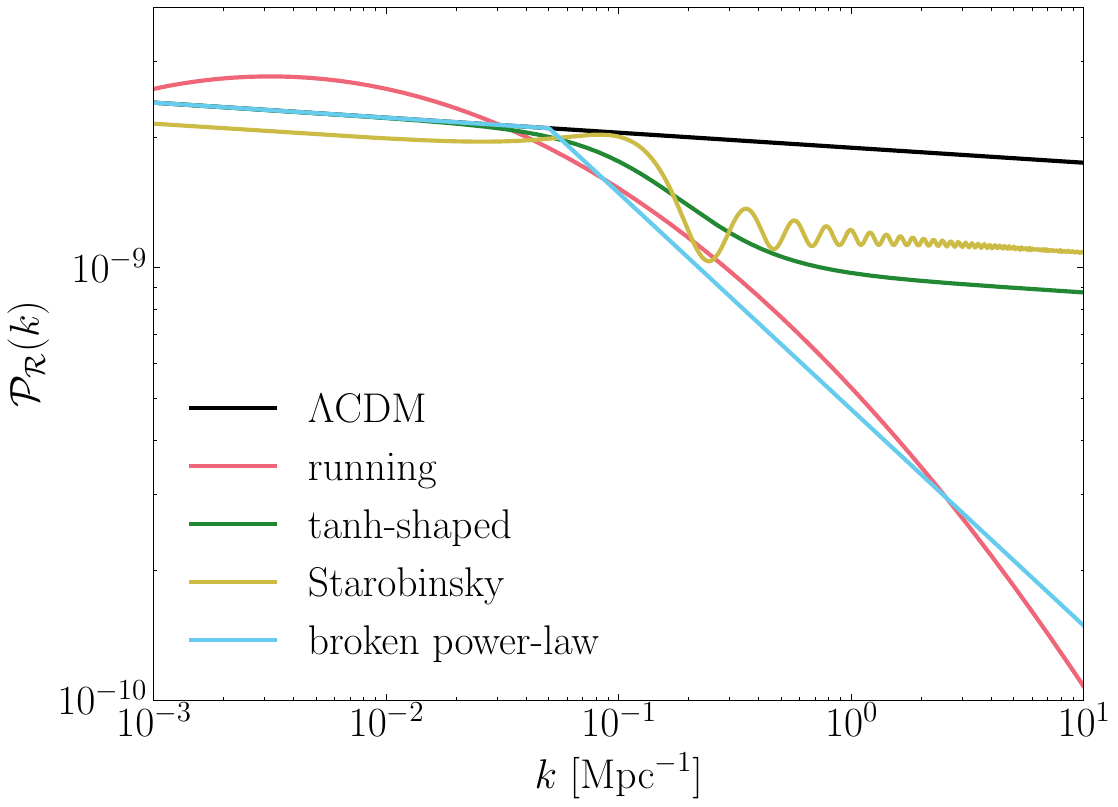}
    \caption{{Modifications in the shape of the primordial power spectrum (PPS) considered
    in this paper (see}
    Section~\ref{subsec:templates} {for details}).
    {For comparison, black line shows the single power-law PPS model, 
    which is commonly assumed in cosmological analyses based on the minimal flat $\Lambda$CDM model with adiabatic initial conditions.}
    }
    \label{fig:PPS_forCS}
\end{figure}
\begin{figure*}
    \includegraphics[width=2\columnwidth]{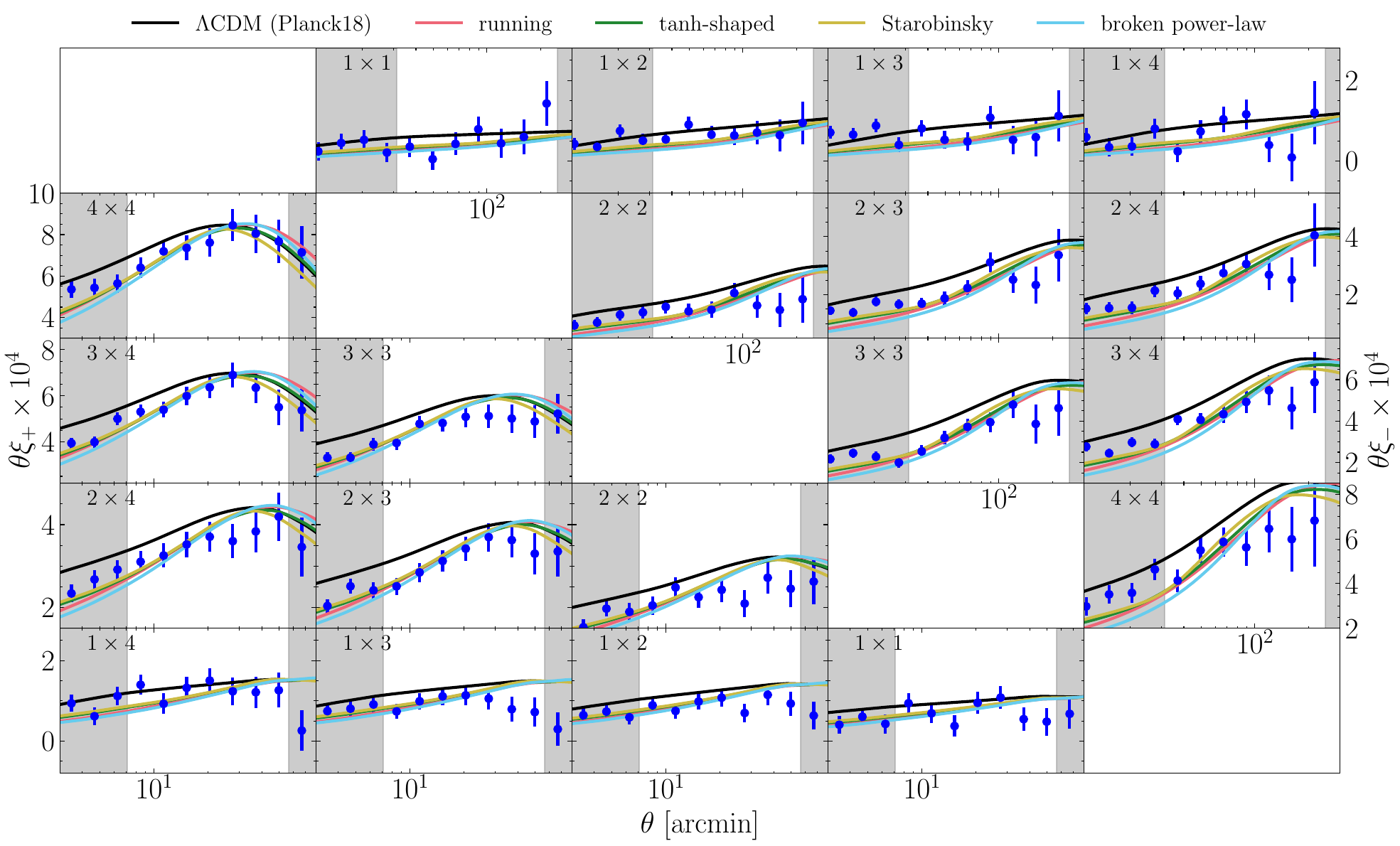}
    \caption{{The model predictions of cosmic}
    shear two-point correlation functions (2PCFs) {{for} the HSC-Y3 data}: the lower-left diagonal panels are for $\xi_+(\theta)$, while 
    the upper-right diagonal panels are for $\xi_-(\theta)$. Different panels show the auto- and cross-2PCFs for galaxies in two tomographic redshift bins; for instance, $\xi_{\pm}$ for ``$3\times 4$'' are the 2PCFs of source galaxies 
    in the 3rd- and 4th- redshift bins. 
    Note that the panels in the same row have the same $y$-range for $\xi_+$ and $\xi_-$ respectively.
    For illustration, 
    we plot $\theta \times \xi_{\pm} \times 10^4$.
    {The different color lines in each panel show the model predictions assuming the modified 
    PPS models in Fig.~\ref{fig:PPS_forCS}, while the} 
    black line shows the model prediction 
    {assuming the single power-law PPS model}. 
    {Note that we here fixed the background cosmological parameters, 
    such as $\Omega_{\rm m}$, to those of the flat $\Lambda$CDM model.}
    For comparison, 
    the data points with error bars denote the measurements from the HSC Year~3 data. The unshaded $x$-axis region denotes the fiducial scale cuts of $\theta$, which are used for the cosmology inference {in this paper, following} 
    \citet{HSC3_cosmicShearReal}.    }
    \label{fig:Effect_on_2PCFs}
\end{figure*}

We consider several parameterized models for the PPS,
denoted as ${\cal P}_{\mathcal{R}}(k)$.
Some of these models have a pivot scale, and we take $k_p = 0.05~{\rm Mpc}^{-1}$ 
throughout this paper.

\subsubsection{{Single power-law model}}
The standard $\Lambda$CDM model {usually} assumes the power-law functional form 
{for PPS.}
We denote {this model}
as ${\cal P}_{\rm st}(k)$:
\begin{align}
    {\cal P}_{\rm st}(k) = A_s \left(\frac{k}{k_p} \right)^{n_s - 1},
\end{align}
where $A_s$ and $n_s$
are the amplitude and spectral tilt parameters of the PPS at 
$k_{p}=0.05~{\rm Mpc}^{-1}$. 
{Note that we use the dimensional-less primordial power spectrum 
${\cal P}(k)$ throughout this paper.}
In Fig.~\ref{fig:PPS_forCS}, the black line shows the PPS assuming the {single}
power-law model with $A_s = 2.1 \times 10^{-9}$ and $n_s= 0.965$, which is
consistent with the constraints from \textit{Planck}-2018 CMB analysis~\citep{cmb_Planck2018_Cosmology}.

\subsubsection{{Running of the spectral index}}
It is a natural extension of the single power-law model to include the higher-order terms of the expansion of $\ln{\cal P}_{\mathcal{R}}$ in $\ln{k}$ as
\begin{align}
    {\cal P}_{\mathcal{R}}(k) = A_s \left(\frac{k}{k_p} \right)^{n(k)} = A_s \left(\frac{k}{k_p} \right)^{n_s - 1 + \frac{\alpha_s}{2}\ln{\left(\frac{k}{k_p} \right)}},
\end{align}
where $\alpha_s = dn/d\ln{k}|_{k=k_p}$ is {a parameter that characterizes the running of the spectral 
index of PPS
(hereafter referred to simply as ``running'').}
Inflationary models {generally predict a} 
non-zero running, {with a typical value of} 
$\alpha_s \simeq -0.001$~\citep{2014PhRvD..89j3527G}.
{The {\it Planck} constraint is given as}
$\alpha_s = -0.0045 \pm 0.0067$ ($68\%$~C.L.)~\citep{Planck2018_Inflation}, and 
{measuring}
the running parameter is one of {the key objectives for upcoming LSS}
surveys~\citep[e.g.][]{Takada:2005si, Euclid2018}. 
{In this paper, we adopt a flat prior of $\alpha_s$
given as}
$\alpha_s \in [-1, 1]$, following the Ref~\cite{Planck2018_Inflation}.
In Fig.~\ref{fig:PPS_forCS}, the red line shows the PPS with the running of $\alpha_s = -0.1$ and $\{A_s, n_s\} = \{1.9 \times 10^{-9}, 0.72\}$. 
Unlike the other templates considered in this paper, {the model with a non-zero running
deviates}
from the single power-law model not only at small scales but also at scales larger than the pivot scale.

\subsubsection{{Tanh-shaped model}}
{Although somewhat artificial, we consider the following PPS model, modified at small scales
following}
\citet{2017PhRvD..95l1302N}:
\begin{align}
    {\cal P}_{\mathcal{R}}(k) &= {\cal P}_{\rm st}(k) \tilde{f}(k), 
\end{align}
where
\begin{align}
    \tilde{f}(k) &\equiv \frac{1+10^{-\alpha_{\rm supp}}}{2} - \frac{1-10^{-\alpha_{\rm supp}}}{2} \tanh{\left(\log \frac{k}{k_{\rm supp}}\right)}.
\end{align}
Here $\alpha_{\rm supp}$ {is a parameter that quantifies}
the amount of the suppression, and 
$k_{\rm supp}$ {is a parameter that defines the wavenumber below which the suppression begins.
Note $\alpha_{\rm supp}=0$ corresponds to $\tilde{f}(k)=1$, i.e. 
no deviation from the single power-law model.
The fudge function $\tilde{f}$ approaches a limit of
$\tilde{f}(k)\rightarrow 1$ at $k\ll k_{\rm supp}$, and}
the PPS {amplitude} is suppressed by a factor of $10^{-\alpha_{\rm supp}}$ 
{at} $k \gtrsim k_{\rm supp}$ compared to the standard power spectrum ${\cal P}_{\rm st}$.
In the {following} analysis we adopt the {flat}
priors {given as}
$\alpha_{\rm supp} \in [0.0, 9.0]$ and $\log k_{\rm supp} \in [-2.0, 0.0]$.
In Fig.~\ref{fig:PPS_forCS}, the green line shows 
{the tanh-shaped model}
with $\alpha_{\rm supp} = 0.3$ and $\log k_{\rm supp} = -0.8$.

\subsubsection{{Starobinsky model}}
A sharp change in the slope of the inflaton potential $V(\phi)$ leads to 
{a} step-like suppression in the primordial power spectrum, 
along with {an}
oscillatory feature \citep{Starobinsky:1992ts,2006PhRvD..74d3518S,2023JCAP...06..018P}.
We adopt the analytic expression, {taken from Eq.~(10) of \citet{Starobinsky:1992ts},}
to model the modified PPS:
\begin{align}
    \label{eq:Starobinsky}
    {\cal P}_{\mathcal{R}}(k) &= {\cal P}_{\rm st}(k) T(k/k_{*}, R_*),
\end{align}     
where
\begin{align}
    T(y, R_*) &\equiv \left[1 - 3(R_* - 1) \frac{1}{y} \left\{ \left(1 - \frac{1}{y^2}\right) 
    \sin{2y} \right.\right.\\ \nonumber 
    &\left.+\frac{2}{y} \cos{2y}\right\} + \frac{9}{2} (R_* - 1)^2 \frac{1}{y^2} 
    \left( 1 + \frac{1}{y^2} \right)
    \\ \nonumber
    &\times \left.\left\{1 + \frac{1}{y^2} + \left(1 - \frac{1}{y^2}\right) \cos{2y} - \frac{2}{y} \sin{2y}
    \right\}\right].
\end{align}
{Here $R_*$ is a parameter that defines the ratio of the inflaton potential 
{slopes} $\mathrm{d}V/\mathrm{d}\phi$ {before and after the sharp change}, and $k_*$ is a parameter that specifies the wavenumber corresponding to the scale at which the sharp change in the potential slope occurs. 
$R_*=1$ corresponds to no deviation from ${\cal P}_{\rm st}$.}
{The limit $y\rightarrow 0$ ($k\ll k_*$) leads to $T(y\rightarrow 0, R_*) = R_*^2 $ and the limit $y\rightarrow \infty$ ($k\gg k_*$) leads to $T(y\rightarrow\infty, R_*)=1$.}
{Therefore, when $R_*>1$, this model predicts a suppression in the PPS amplitude at small scales.}
In the {following} analysis we adopt flat priors given as $R_{*} \in [0.01, 2.0]$ and $\log k_{*} \in [-2.5, 0.0]$.
In Fig.~\ref{fig:PPS_forCS}, the yellow line shows the PPS of the Starobinsky model with $R_* = 1.2$ and $\log k_* = 0.067$.
{Note that} we set $A_s = 1.3 \times 10^{-9}$ in the plot, 
{and the effective spectral index at {$k\ll k_*$ and} $k\gg k_*$ remains the same as $n_s$ for the 
single power-law
PPS}.

\subsubsection{{Broken power-law model}}
We also consider a broken power-law {model}
for the PPS~\citep[e.g.][]{2024JCAP...12..038H}, parametrized as
\begin{align}
    {
    {\cal P}_{\mathcal{R}}(k) = 
    \begin{cases}
    {\cal P}_{\rm st}(k), \hspace{2em} & (k < k_{\rm break})\\
    {\cal P}_{\rm st}(k_{\rm break}) \left(\frac{k}{k_{\rm break}} \right)^{n_{s,2} - 1},  & (k_{\rm break} \leq k)
    \end{cases}
    }
\end{align}
{where $k_{\rm break}$ is the breakpoint wavenumber that separates the two power-law regimes, 
and $n_{s,2}$ is a parameter that defines {the spectral tilt in} the second power-law regime at small scales.}
In the {following}
analysis we adopt {flat}
priors {given as}
$\log k_{\rm break} \in [-2.5, 0.0]$ and $n_{s,2} \in [0.5, 1.1]$.
In Fig.~\ref{fig:PPS_forCS}, the cyan line shows 
the broken power-law model with $n_{s,2} = 0.5$ and $\log k_{\rm break} = -1.3$.
{We will also introduce an alternative parametrization for $n_{s,2}$ 
as
{$n_{s,2}-1 \equiv n_s-1-\Delta n_s$, 
where $\Delta n_s\equiv n_s-n_{s,2}$.}
In this case, $\Delta n_s=0$ corresponds to no deviation from 
${\cal P}_{\rm st}$.}

\subsubsection{{Multiple broken power-law model}}
Alternatively, following Ref~\cite{Planck2015_Inflation}, we also model the PPS 
as piecewise linear in the 
{$\ln {\cal P}_{\mathcal{R}}$-$\ln k$}
plane, {with} 
a number of knots, $N_{\rm knots}$, that is allowed to vary. 
We construct {this} 
model by placing $N_{\rm knots}$ knots $\{(k_i,{\cal P}_i): i = 1,...,N_{\rm knots}\}$ in the $(\ln k, \ln {\cal P}_{\mathcal{R}})$ plane.
{We construct the functional form of}
${\cal P}_{\mathcal{R}}(k)$ 
{using}
linear interpolation in log-log space between adjacent points, 
where we fix the end-points for the wavenumber as $k_1 = 10^{-4}~{\rm Mpc}^{-1}$ and $k_{N_{\rm knots}} = 1~{\rm Mpc}^{-1}$. 
{Therefore, we have}
$2N_{\rm knots}-2$ parameters {to specify this PPS model.}

{Note that $N_{\rm knots}=2$ corresponds to the single power-law PPS model, 
and $N_{\rm knots}=3$ corresponds to the broken power-law model.
Therefore, when $N_{\rm knots}\ge 4$, 
this model gives a generalization of the broken power-law model 
with multiple breaking wavenumber points.
For this reason, we refer to this model as the ``multiple broken power-law model''.}
In this {paper,}
we consider $N_{\rm knots} = 4$, and 
{we adopt flat priors for}
$\{(k_i,{\cal P}_i)\}$ {as given}
in Table~\ref{tab:parameters_free-form}.

\subsection{The impact of the modified PPS on cosmic shear 2PCFs}
\label{subsec:cosmic shear}

{In this subsection, we study how small-scale modifications in the PPS shape affect
the cosmic shear two-point correlation functions (2PCFs), which are standard summary statistics used in cosmological analyses.}

With the flat-sky approximation, the cosmic shear 
2PCFs
can be expressed 
in terms of the $E$- and $B$-mode angular power spectra $C^{E/B}(\ell)$ via the Hankel transform:
\begin{align}
\label{eq:model_hankel}
\xi^{ij}_{+/-}(\theta) = 
\int\!\frac{\ell \mathrm{d}\ell}{2\pi} \,
        \left[ C^{E;ij}(\ell) \pm C^{B;ij}(\ell) \right]J_{0/4}(\theta \ell) ,
\end{align}
where $J_{0/4}$ are the $0$-th/$4$th-order Bessel functions of the first kind and 
the transformation of $\xi_+ (\xi_-)$
uses $J_0 (J_4)$, respectively.
In our analysis, we use 
{\textsc{FFTLog}}
\citep{fftlog2020} to perform 
the Hankel transform.
The superscripts ``$ij$'' denote tomographic 
{redshift}
bins; e.g. ``$ij$'' means the 2PCF or power spectrum obtained using 
source galaxies in the $i$-th and $j$-th tomographic redshift bins. 

The $E$- {and $B$-}mode angular power spectra in Eq.~(\ref{eq:model_hankel}) are given as
\begin{align}
C^{E;ij}(\ell)&=\int^{\chi_H}_0\!\mathrm{d}\chi\frac{q_i(\chi)q_j(\chi)}{\chi^2}
P_{\rm m}\!\!\left(k=\frac{\ell+1/2}{\chi};z(\chi)\right) 
\label{eq:cl_EE}
\end{align}
and
\begin{align}
C^{B;ij}(\ell)&= 0,
\label{eq:cl_BB}
\end{align}
where $P_{\rm m}(k;z)$ is the nonlinear
matter power spectrum at redshift $z$, 
$\chi(z)$ is the radial comoving distance up to redshift $z$, and
$\chi_H$ is the distance to the horizon.
The lensing efficiency function for source galaxies in the $i$-th redshift bin, $q_i(\chi)$, 
is
defined as
\begin{align}
q_i(\chi)=\frac{3}{2}\frac{\Omega_{\rm m}H_0^2}{c^2}\frac{\chi}{a(\chi)}\int_\chi^{\chi_H}\!\mathrm{d}\chi'~n_i(\chi')\frac{\chi'-\chi}{\chi'},
\end{align}
where $\Omega_{\rm m}$ is the present-day density parameter of matter, $H_0$ is the present-day Hubble constant ($H_0=100h~{\rm km}~{\rm s}^{-1}~{\rm Mpc}^{-1}$), $a$ is the scale factor, $n_i(\chi)$ is the normalized redshift distribution of galaxies in the $i$-th source redshift bin, {and $c$ is the speed of light.}

{As can be found Eq.~(\ref{eq:cl_EE}), we need to model the nonlinear matter power spectrum,
 $P_{\rm m}(k;z)$, 
for an input cosmological model. In this paper, we use the \textsc{HMCode16}~\citep{halofit_mead16},
which models the nonlinear power spectrum by applying a nonlinear mapping to the input linear matter power spectrum at a given redshift for a given cosmological model.
Here the input linear matter power spectrum is given by the product of 
the primordial power spectrum and the transfer function.
Assuming the adiabatic initial conditions, we use the \textsc{CAMB}~\citep{CAMB2000} code 
to model the transfer function for a given flat $\Lambda$CDM model, based on cosmological linear perturbation theory, which is a well-established and accurate framework without ambiguity
\citep{1984PThPS..78....1K} \citep[also see][]{2020moco.book.....D}.
For the input PPS, we consider various models, as given in Sections~\ref{subsec:templates}. Note that}
the \textsc{HMCode16} includes the parameters to model the baryonic effect.
Here we assume \textsc{HMCode16} can still {model the nonlinear matter power spectrum 
even when the input PPS includes shape modifications,
as considered in this paper.}
{Validating our method requires N-body simulations using the input PPS, but we defer this to future work.}

We consider the fiducial scale cuts as in \citet{HSC3_cosmicShearReal} \citep[also see][]{Terasawa24};
{we use the data vector of $\xi_+$ in the range}
$\theta \in [7.1, 56.6]$~arcmin and the data of $\xi_-$ in the range 
 $\theta \in [31.2, 247.8]$~arcmin, respectively. 
The wavenumbers of the matter power spectrum 
contributing to $\xi_{\pm}$ at 
these scales are $k \sim [10^{-2}, 1]~h~{\rm Mpc}^{-1}$ (see Fig. 1 of \citet{Terasawa24}).

In Fig.~\ref{fig:Effect_on_2PCFs}, {we show how the different PPS models
 give varying 
predictions for the cosmic shear 2PCFs, compared to the 2PCFs measured from the Subaru HSC-Y3 data
\citep{HSC3_cosmicShearReal}. Here we assumed the best-fit flat $\Lambda$CDM model that is consistent 
with the {\it Planck} CMB data \cite{Planck2018_Inflation}, for the background cosmology. 
The black solid line is the prediction based on the standard single power-law PPS, using the best-fit values of $A_s$ and $n_s$ consistent with the {\it Planck} $\Lambda$CDM cosmology, as shown in Fig.~\ref{fig:PPS_forCS}. The prediction shows higher amplitudes than the HSC-Y3 result, reflecting the $S_8$
tension between the $\Lambda$CDM models consistent with 
the {\it Planck} data and the HSC-Y3 data~\cite{HSC3_cosmicShearReal,Terasawa24}.}
Other curves in Fig.~\ref{fig:Effect_on_2PCFs} are the cosmic shear 2PCFs {computed using the modified PPS models}
in Fig.~\ref{fig:PPS_forCS}. 
{If we consider PPS modifications where the modified PPS has smaller amplitudes at small scales,
$k\gtrsim 0.1~h$Mpc$^{-1}$, as shown in Fig.~\ref{fig:PPS_forCS}, the resulting cosmic shear 2PCFs exhibit reduced amplitudes on the angular scales probed by the HSC-Y3 data, thereby alleviating the 
$S_8$ tension. Thus, we expect that combining the HSC-Y3 cosmic shear data with other cosmological probes, such as the {\it Planck} CMB data and CMB lensing data, will enable us to constrain the PPS shape. In the following, we present a more quantitative analysis of the PPS constraints.}

\section{Analysis Method}
\label{sec:analysis}

\subsection{Data}
\label{sec:data}

\begin{table}
\caption{Model parameters and priors used in our 
cosmological parameter inference {using the different PPS models (see Fig.~\ref{fig:PPS_forCS}).}
The label ${\mathcal U}(a,b)$ denotes a uniform flat prior
between $a$ and $b$, and ${\mathcal N}(\mu, \sigma)$ denotes a normal
distribution with mean $\mu$ and width $\sigma$.
{All our analyses {except for the one with the multiple broken power-law model} use the same six cosmological parameters, which are listed  
in the upper table enclosed by the double horizontal lines.
Each of the lower tables denotes ``additional'' parameters used to characterize the modified PPS models.
Note that we use units ${\rm Mpc}^{-1}$ (without 
$h$) for the wavenumber of 
PPS.} 
We fix the nuisance parameters associated with the cosmic shear 2PCFs to the MAP values in \citet{HSC3_cosmicShearReal} (see text for details).
}
\label{tab:parameters}
\setlength{\tabcolsep}{20pt}
\begin{center}
\begin{tabular}{ll}  \hline\hline
Parameter & Prior \\ \hline
\multicolumn{2}{l}{\hspace{-1em}\bf Cosmological parameters}
\\
$\ln 10^{10}A_\mathrm{s}$   & ${\cal U}(1.61, 4.0)$\\
$n_\mathrm{s}$                      & ${\cal U}(0.9, 1.1)$\\
$h$                               & ${\cal U}(0.2, 1.0)$\\
$\omega_\mathrm{b}$                 & ${\cal U}(0.005, 0.1)$\\
$\omega_\mathrm{c}$                 & ${\cal U}(0.001, 0.99)$\\
$\tau$                    & ${\cal N}(0.0506, 0.0086)$ \\
\hline
\hline
\multicolumn{2}{l}{\hspace{-1em}\bf PPS template parameters} \\
\hline
\multicolumn{2}{l}{\hspace{-1em}\bf running} \\
$\alpha_\mathrm{s}$                 & ${\cal U}(-1.0, 1.0)$\\
\hline
\multicolumn{2}{l}{\hspace{-1em}\bf tanh-shaped  model} \\
$\log_{10}k_{\rm supp}~[{\rm Mpc}^{-1}]$                 & ${\cal U}(-2.0, 0.0)$\\
$\alpha_{\rm supp}$                 & ${\cal U}(0.0, 9.0)$\\
\hline
\multicolumn{2}{l}{\hspace{-1em}\bf Starobinsky model} \\
$\log_{10}k_{*}~[{\rm Mpc}^{-1}]$                 & ${\cal U}(-2.5, 0.0)$\\
$R_{*}$                 & ${\cal U}(0.01, 2.0)$\\
\hline
\multicolumn{2}{l}{\hspace{-1em}\bf broken power-law model} \\
$\log_{10}k_{\rm break}~[{\rm Mpc}^{-1}]$                 & ${\cal U}(-2.5, 0.0)$\\
$n_{s,2}$                 & ${\cal U}(0.5, 1.1)$\\
\hline
\hline
\end{tabular}
\end{center}
\end{table}

\begin{table}
\caption{{Model parameters used to characterize the multiple broken power-law 
PPS model (see Section~\ref{subsec:templates}).
In this paper, we consider $N_{\rm knots}=4$. In this case 
we have 6 parameters: {4 amplitude parameters at the end-points and the breakpoints and 2 breakpoints ($k_2, k_3$).}
Note we fix the end-points 
{for the wavenumber as}
$k_1=10^{-4}$
and $k_4=1~{\rm Mpc}^{-1}$.}
}
\label{tab:parameters_free-form}
\setlength{\tabcolsep}{12pt}
\begin{center}
\begin{tabular}{ll}  \hline\hline
Parameter & Prior \\ \hline
$\ln{(10^{10}{\cal P}_{1,\cdots, N_{\rm knots}})}$                 & ${\cal U}(2,4)$\\
$\log_{10}k_2 <,\cdots, < \log_{10} k_{N_{\rm knots}-1}~[{\rm Mpc}^{-1}]$   & ${\cal U}(-4,0)$\\
$\log_{10} k_{1}~[{\rm Mpc}^{-1}]$   & $-4~$ (fixed)\\
$\log_{10}k_{N_{\rm knots}}~[{\rm Mpc}^{-1}]$   & $0~$ (fixed)\\
\hline
\hline
\end{tabular}
\end{center}
\end{table}

{We perform a cosmological analysis by combining different cosmological datasets 
to estimate both cosmological parameters and the PPS parameters. We use}
the HSC-Y3 cosmic shear 2PCFs~\citep{HSC3_cosmicShearReal},
the \textit{Planck}-2018 CMB power spectra~\citep{cmb_Planck2018_Cosmology},
the ACT DR6 CMB lensing data~\citep{ACTDR6_Qu, ACTDR6_Madhavacheril},
and the DESI Y1 BAO data~\citep{DESI_BAO}. 
The datasets are summarized as {follows.}
\begin{enumerate}
    \item \textbf{HSC-Y3}: The HSC-Y3 {weak lensing measurements
    were based on the catalog of 25~million galaxies over about 416~deg$^2$}
        (with $n_\text{eff}$$\sim$$15~\mathrm{arcmin}^{-2}$) 
        \citep{HSC3_catalog_Li2021}. 
        {In this paper, we use the data vector containing}
        the cosmic shear 2PCFs measured from the HSC-Y3 data in \citet{HSC3_cosmicShearReal}.
        We employ the fiducial scale-cut in \citet{HSC3_cosmicShearReal} ($\theta \in [7.1, 56.6]$ arcmin for $\xi_+$ and $\theta \in [31.2, 247.8]$ arcmin for $\xi_-$), which corresponds to $k \sim [10^{-2}, 1]~{\rm Mpc}^{-1}$~\citep{Terasawa24}. 
    \item \textbf{Planck-2018:} {We use} 
    the third data release (PR3) from the
        \textit{Planck} 
        experiment
        \citep{cmb_Planck2018_Cosmology}. 
        We {use the data vector containing}
        the primary $TT$, $TE$, and $EE$ {power spectra in the multipole range}
        $30< \ell < 2508$, which corresponds to $k \sim [2\times 10^{-3}, 0.2]~{\rm Mpc}^{-1}$.
         We do not incorporate the low-$\ell$ ($2 <
        \ell < 30$) likelihood. Instead, we {employ a Gaussian prior}
        on the optical depth $\tau$ inferred from the
        low-$\ell$ $EE$ data: ${\cal N}(0.0506, 0.0086)$.
    \item \textbf{ACT DR6:} 
    We use the CMB lensing power spectrum measured from ACT DR6 \citep{ACTDR6_Qu, ACTDR6_Madhavacheril}.
     {We {use the data vector defined within}
     the baseline multipole range in \cite{ACTDR6_Qu, ACTDR6_Madhavacheril}, $40 < L
     < 763$}, which corresponds to $k \sim [4\times 10^{-3}, 0.2]~{\rm Mpc}^{-1}$.
    \item \textbf{DESI Y1 BAO:} We include {the information of}
    baryon acoustic
        oscillation (BAO) measurements from the DESI Year~1 galaxy sample \citep{DESI_BAO}. We
        use {the BAO information for all galaxy samples,}
       ``BGS", ``LRG1", ``LRG2", ``LRG3+ELG1", ``ELG2", ``QSO", and ``Lya QSO".
        {The BAO information help constrain 
        the cosmological parameters of $\Lambda$CDM model, such as $\Omega_{\rm m}$, when
         combined with the {\it Planck} CMB information.} 
\end{enumerate}

\subsection{Parameter inference}
Our analysis uses a set of parameters and priors summarized in Table~\ref{tab:parameters}.
The parameters include six cosmological parameters, denoted by $\mathbb{C}=\left\{\ln 10^{10} A_s, n_s,
h, \omega_{\rm b}, \omega_{\rm c}, \tau \right\}$,
for flat $\Lambda$CDM cosmologies.
$h$ is the Hubble parameter, and 
$\omega_{\rm b}(\equiv \Omega_{\rm b}h^2)$ is the physical density parameter of baryon. 
The physical density parameter of CDM is given as $\omega_{\rm c}(\equiv \Omega_{\rm c}h^2)=\Omega_{\rm m}h^2-\omega_b-\omega_\nu$,
where $\omega_\nu$ is the physical density parameter of massive neutrinos and we adopted the fixed total neutrino mass, $M_\nu=0.06~{\rm eV}$. 
The parameters specifying the PPS are described in Table~\ref{tab:parameters} and Section~\ref{subsec:templates}.
For the analysis with the multiple broken power-law model,
we adopt the priors on those PPS parameters $\{(k_i,{\cal P}_i)\}$ as in Table~\ref{tab:parameters_free-form} and vary other four cosmological parameters, $\left\{
h, \omega_{\rm b}, \omega_{\rm c}, \tau \right\}$.

We use \textsc{CAMB}~\citep{CAMB2000}, which {allows us to input the modified PPS model, }
to calculate the primary CMB power spectra and CMB lensing power spectrum {assuming 
the adiabatic initial conditions}.
We use the CMB and CMB lensing likelihood where the associated nuisance parameters are already marginalized over. We fix the nuisance parameters for the cosmic shear 2PCFs, including the parameters for the baryonic feedback, intrinsic alignment, photo-$z$ systematics, to the maximum a posteriori (MAP) values of \citet{HSC3_cosmicShearReal}. 
We assume the observations are independent and ignore cross-covariance between them
{so that our total likelihood $\mathcal{L}$ is given by the product of the likelihoods for each dataset.}

We use \textsc{Multinest}~\citep{multinest_Feroz2009} to sample the posterior distribution.
Throughout this paper, we report the 1D marginalized {\em mode} and its asymmetric 68\% credible intervals,
together with the MAP estimated as the maximum of the posterior in the chain
\citep[see Eq.~1 in][]{HSC1_2x2pt_Miyatake2022}. For the 2D marginalized posterior, 
we report the mode and the 68\% and 95\% credible intervals. 
We use \textsc{GetDist} \citep{GetDist2019} for the plotting.

To compare the different models for the PPS and the $\Lambda$CDM model, we use goodness-of-fit $\chi^2 = -2\ln \mathcal{L}^{\rm MAP}$\footnote{{Here, we omit the constant term corresponding to the likelihood normalization, as it cancels out in $\Delta\chi^2$.}}.
In practice we compare the difference in $\chi^2$ divided by the number of the additional parameters $\Delta p$, namely $\Delta \chi^2/\Delta p$, between the modified PPS model and the $\Lambda$CDM model.
{If}
$\Delta \chi^2$ follows the $\chi^2$ distribution with $\Delta p$ degrees of freedom, its 
{expected}
value is {typically} $\Delta \chi^2/\Delta p \simeq 1$. {Therefore,}
$|\Delta \chi^2/\Delta p| > 1 (< 1)$ means {a}
better (worse) fit, compared to  
the number of the additional parameters.
{We will use $|\Delta\chi^2/\Delta p|$ as an indicator to assess whether the modified PPS model provides a better fit compared to the standard single power-law PPS model.}
We also use the Bayes factor $R = {\mathcal{Z}/\mathcal{Z}_{\rm st}}$,
where $\mathcal{Z}$ is the Bayesian evidence obtained from \textsc{Multinest} sampler and $\mathcal{Z}_{\rm st}$ denotes the Bayesian evidence for the standard {single power-law PPS model.}

\section{Results}
\label{sec:results}

\begin{table}
     \centering
      \caption{{A metric for evaluating model fit improvement with each of the modified PPS models,
      compared to the single power-law PPS model. Here we include all the HSC-Y3 cosmic shear data, 
      {\it Planck} CMB, ACT DR6 CMB lensing and DESI-Y1 BAO data in the parameter inference.
      $\Delta \chi^2/\Delta p$ is the improvement in the $\chi^2$ value for the MAP model, where
      $\Delta p$ is the number of the additional parameters to model the modified PPS
      compared to the single power-law PPS model (see Table~\ref{tab:parameters}).
      $\ln R$ is the logarithmic ratio of the {Bayesian evidence}.}
     {The last column denotes the {strength}
     of evidence in favor of {the single power-law PPS model}
     based on the Jeffreys' scale~\citep{Jeffreys}. 
     }
     }
     \begin{tabular}{c|c|c|l} \hline \hline
        model  &   $\Delta \chi^2 / \Delta p$ & $\ln R$ & Grades of evidence \\ \hline\hline
      running & $-0.03$ & $-4.4$ & Very strong \\ \hline
     $\tanh$-shaped & $-0.16$ & $-5.5$ & Decisive
  \\ \hline
      Starobinsky & $-0.6$ & $-2.3$ & Strong \\ \hline
    broken power-law & $-0.5$ & $-1.5$ & Substantial \\ \hline
   $N_{\rm knot} = 4$ & $-0.5$ & $-1.4$ & Substantial \\ \hline
      \hline
     \end{tabular}
    
     \label{tab:goodness-of-fit}
 \end{table}

\begin{figure}
    \includegraphics[width=\columnwidth]{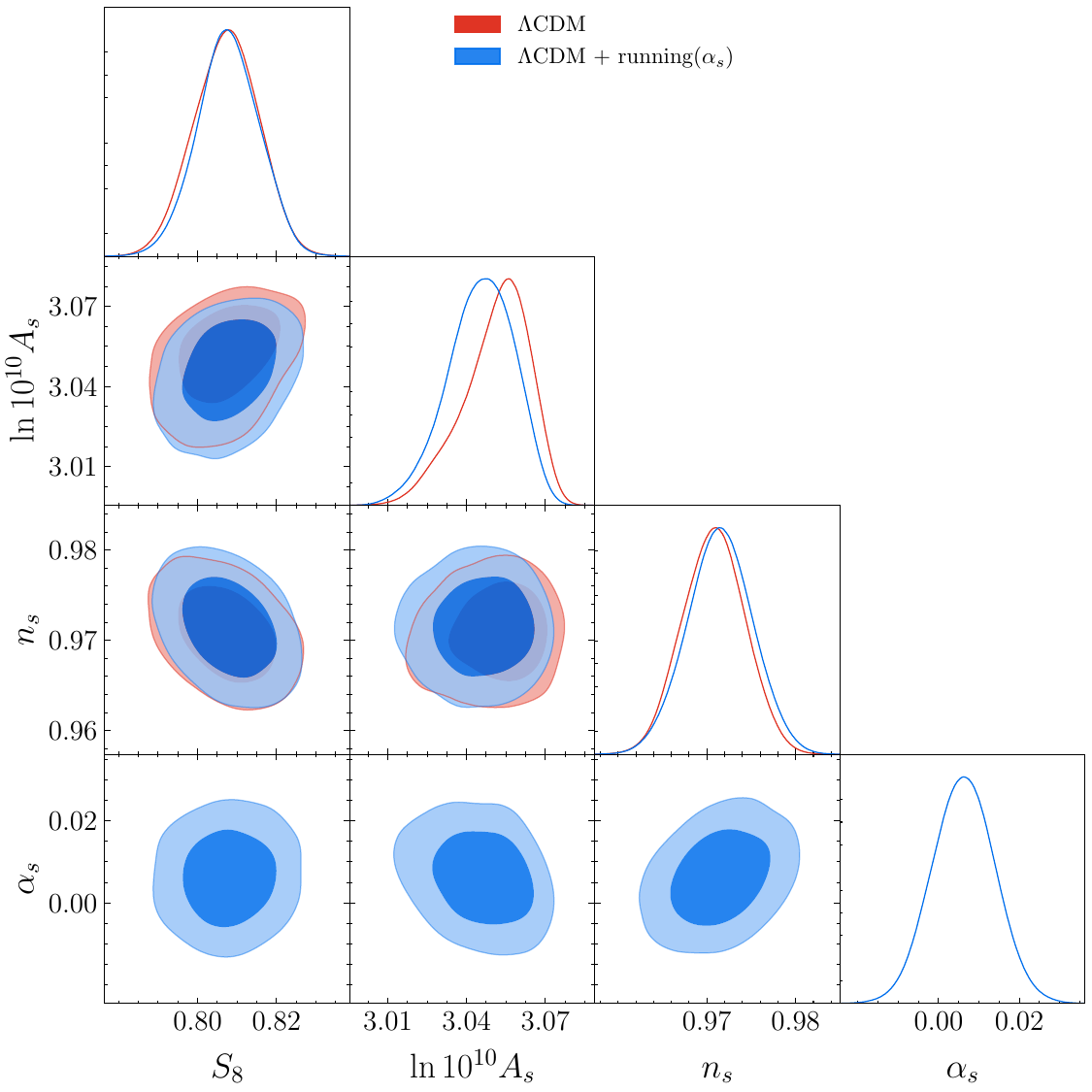}
    \caption{The 1D and 2D posteriors obtained {from the parameter inference 
    combining the HSC-Y3 cosmic shear data, {\it Planck} CMB data, ACT DR6 CMB lensing data,
    and DESI-Y1 BAO data (see Section~\ref{sec:data}).
    The blue contours show the posteriors assuming the PPS model with a running spectral index, while
    the red contours show the posteriors assuming the single power-law PPS model, denoted 
    as ``$\Lambda$CDM''.
    The results include marginalization over other parameters.}
}
    \label{fig:2D_nrun}
\end{figure}

\begin{figure}
    \includegraphics[width=\columnwidth]{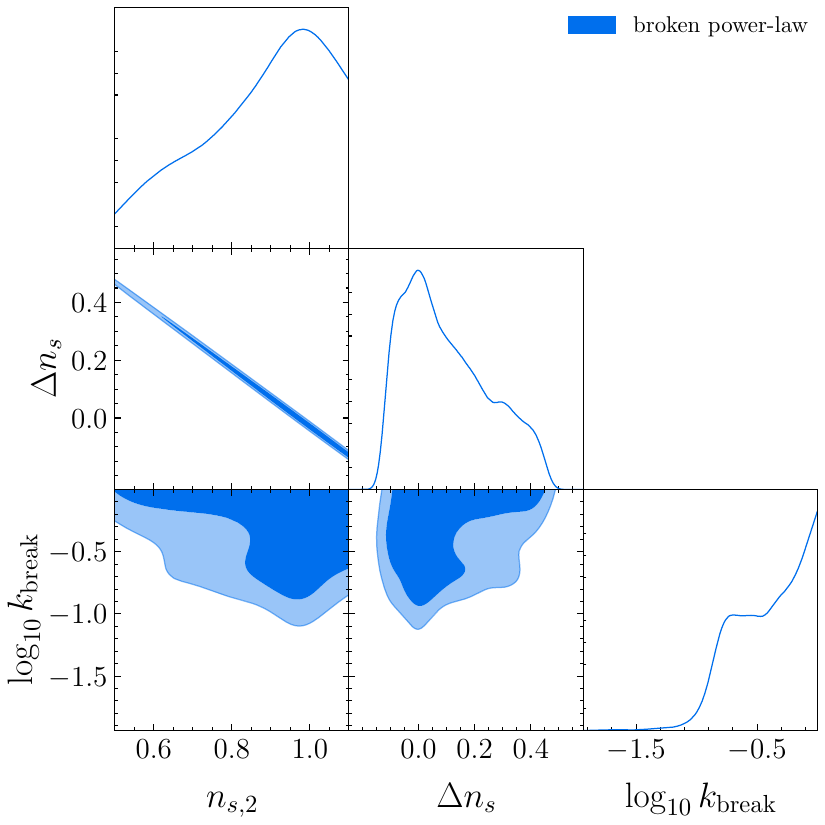}
    \caption{{Similarly to Fig.~\ref{fig:2D_nrun}, the blue contours show the posteriors
    for the broken power-law model.}
{Note that $\Delta n_s=0$ corresponds to the single power-law PPS model.
Here we show the posteriors for the parameters that characterize the PPS. 
The posteriors for other cosmological parameters remain largely unchanged compared to the results 
obtained assuming the single power-law PPS model.
}
}
    \label{fig:2D_BPL}
\end{figure}

\begin{figure}
    \includegraphics[width=\columnwidth]{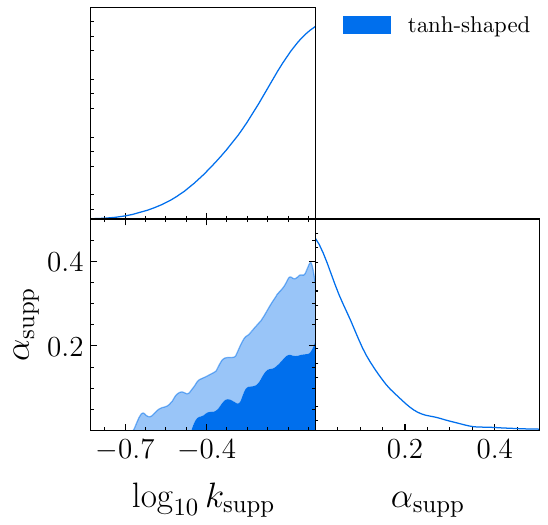}
    \caption{{Similarly to Fig.~\ref{fig:2D_nrun}, the blue contours show the posteriors
    for the tanh-shaped model.} 
{Note that $\alpha_{\rm supp}=0$ corresponds to the single power-law PPS model.
Here we show the posteriors for the parameters that characterize the PPS. 
The posteriors for other cosmological parameters remain largely unchanged compared to the results 
obtained assuming the single power-law PPS model.
}
    }
    \label{fig:2D_tanh}
\end{figure}

\begin{figure}
    \includegraphics[width=\columnwidth]{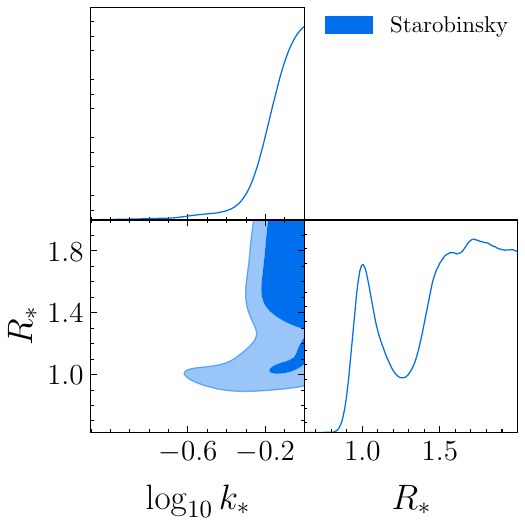}  
    \caption{{Similarly to Fig.~\ref{fig:2D_nrun}, the blue contours show the posteriors
    for the Starobinsky model.} 
{Note that $R_*=1$ corresponds to the single power-law PPS model.
Here we show the posteriors for the parameters that characterize the PPS. 
The posteriors for other cosmological parameters remain largely unchanged compared to the results 
obtained assuming the single power-law PPS model.
}
    }
    \label{fig:2D_SB}
\end{figure}

Fig.~\ref{fig:2D_nrun} shows the posterior distributions obtained from the analysis assuming 
{the single power-law PPS}
(red contours) and {the} analysis {using the PPS model with the running spectral index
$\alpha_s$}
(blue contours).
{We often refer to the minimum single-power law model as the (standard) ``$\Lambda$CDM'' model.}
The two contours are almost indistinguishable except for $\ln 10^{10} A_s$, and 
{the posterior of $\alpha_s$ is}
consistent with $\alpha_s = 0$. 
As noted in Table~\ref{tab:goodness-of-fit}, the inclusion of the running parameter decreases the $\chi^2$ only by $-0.03$, and {the logarithmic} Bayes Factor is as small as $-4.4$.
Hence we conclude there is no preference for the running PPS model
over the single power-law PPS model.
{We note that Refs.~\citep{2020JCAP...04..038P,2025PhRvR...7a2018R} claimed 
that the running-index model with $\alpha_s\sim -0.01$ alleviates a possible tension
between the CMB and Ly-$\alpha$ measurements. However, our combined analysis 
of the HSC cosmic shear and the {\it Planck} CMB data does not 
{prefer} such a model with a negative
$\alpha_s$, while $\alpha_s\sim -0.01$ is acceptable as found in Fig.~\ref{fig:2D_nrun}.}

{Similarly to Fig.~\ref{fig:2D_nrun}, the blue contours in} Figs.~\ref{fig:2D_BPL},~\ref{fig:2D_tanh} and \ref{fig:2D_SB} {show the posteriors for the parameters that characterize the 
modified PPS, for each of {the other three} PPS models shown in Fig.~\ref{fig:PPS_forCS}, where the model parameters are allowed to vary in the inference. For illustrative purpose, 
we show the posteriors only for the PPS parameters, as other cosmological parameters, such as 
$\omega_{\rm b}$, remain nearly unchanged from those of the single power-law model.}
For the broken power-law model, we plot $\Delta n_s \equiv n_s - n_{s,2}$. The posterior is consistent with the single power-law model ($\Delta n_s = 0$). 
{For the tanh-shaped and Starobinsky PPS models,}
 the posteriors are also consistent with their single power-law (the standard $\Lambda$CDM) limit, given by
 $\alpha_{\rm supp} =0$ and $R_* = 1$.

As summarized in Table~\ref{tab:goodness-of-fit}, none of 
{the modified PPS models}
give a positive $\ln R$ or $\Delta \chi^2 / \Delta p < -1$. 
{In particular,}
the 
{running-index and tanh-shaped models are not supported} in terms of the Jeffreys' scale~\citep{Jeffreys}. 
{Therefore,}
we conclude that {the single power-law PPS model is}
preferred over {the modified PPS models considered in this work.}
\begin{figure*}
\includegraphics[width=\columnwidth]{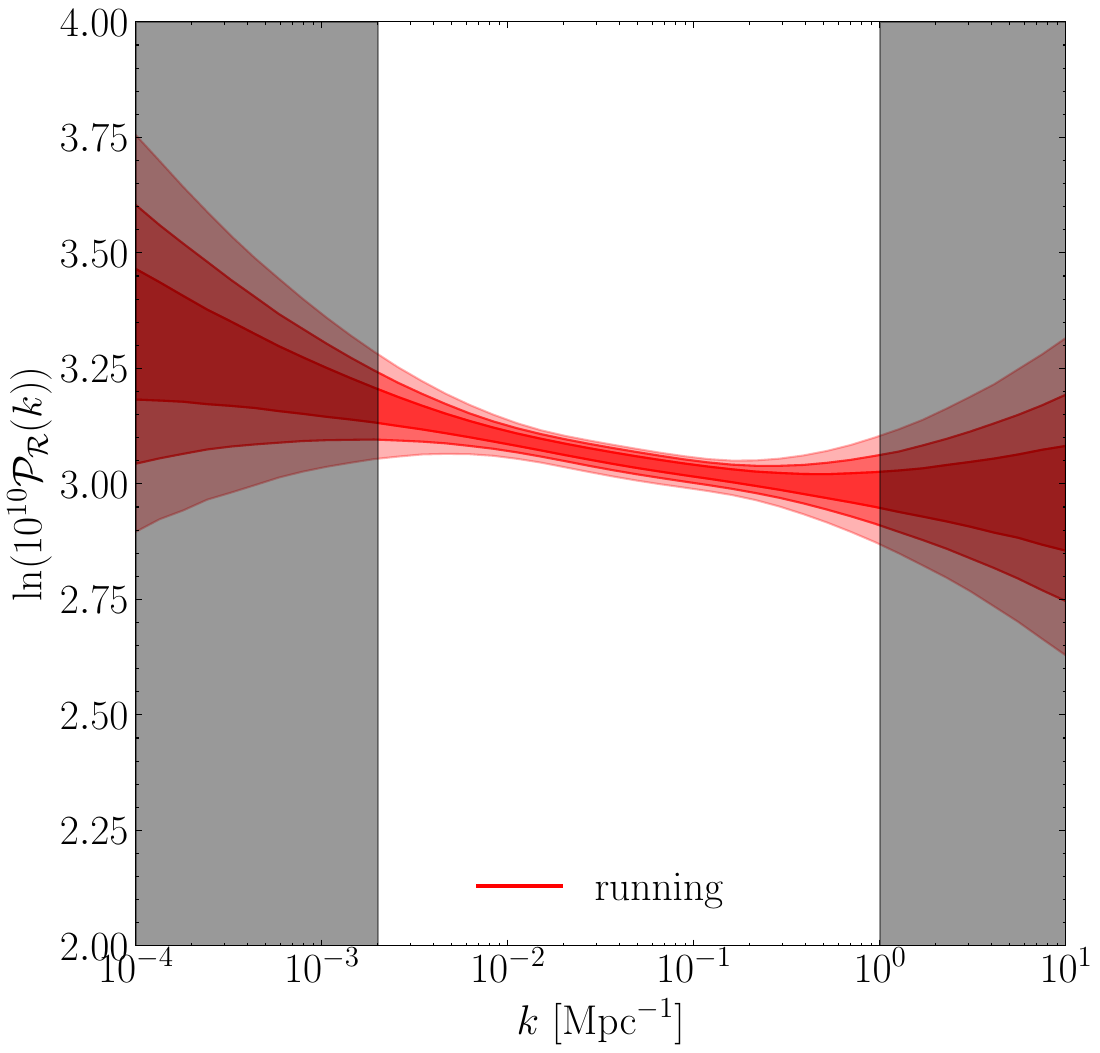}
    \includegraphics[width=\columnwidth]{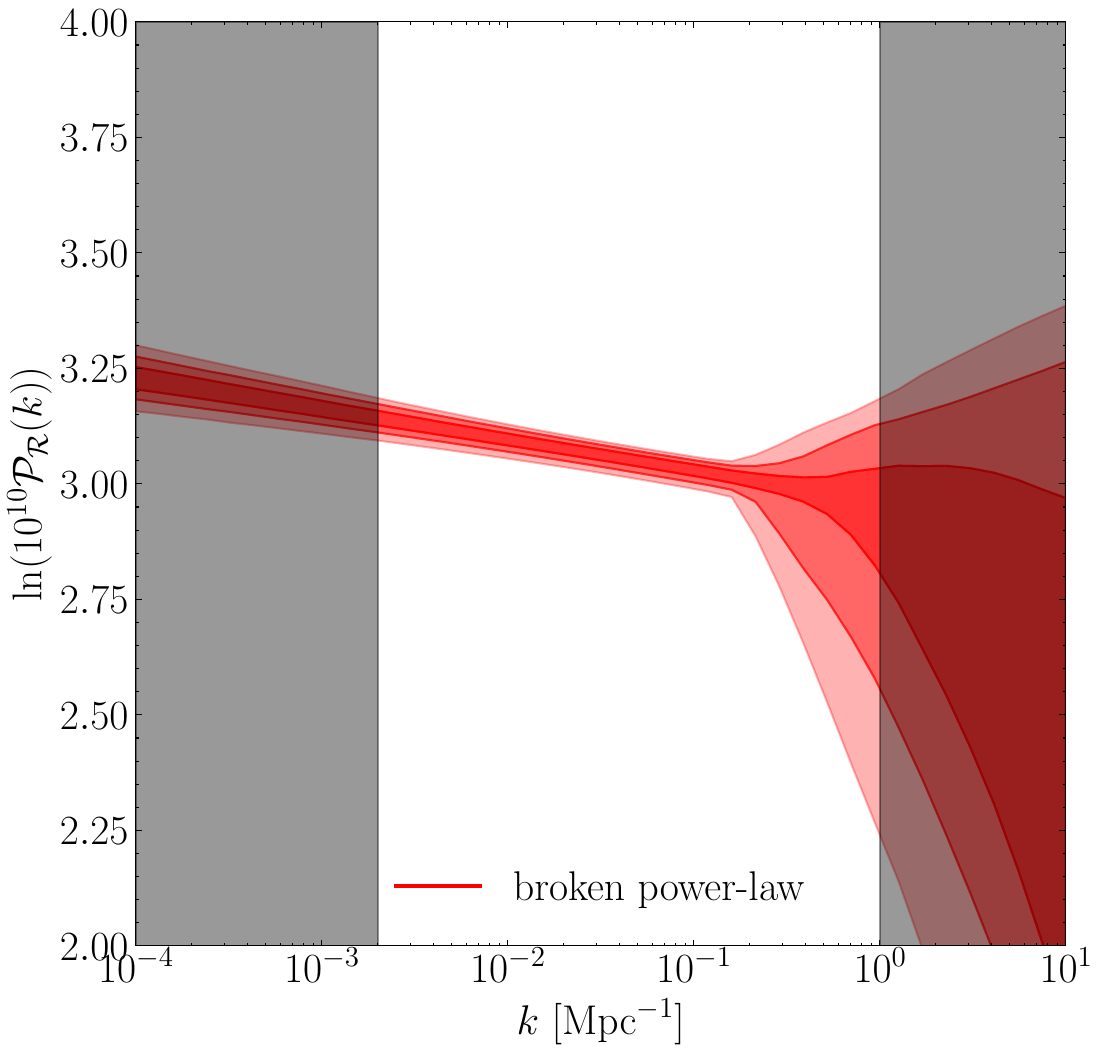}
    \includegraphics[width=\columnwidth]{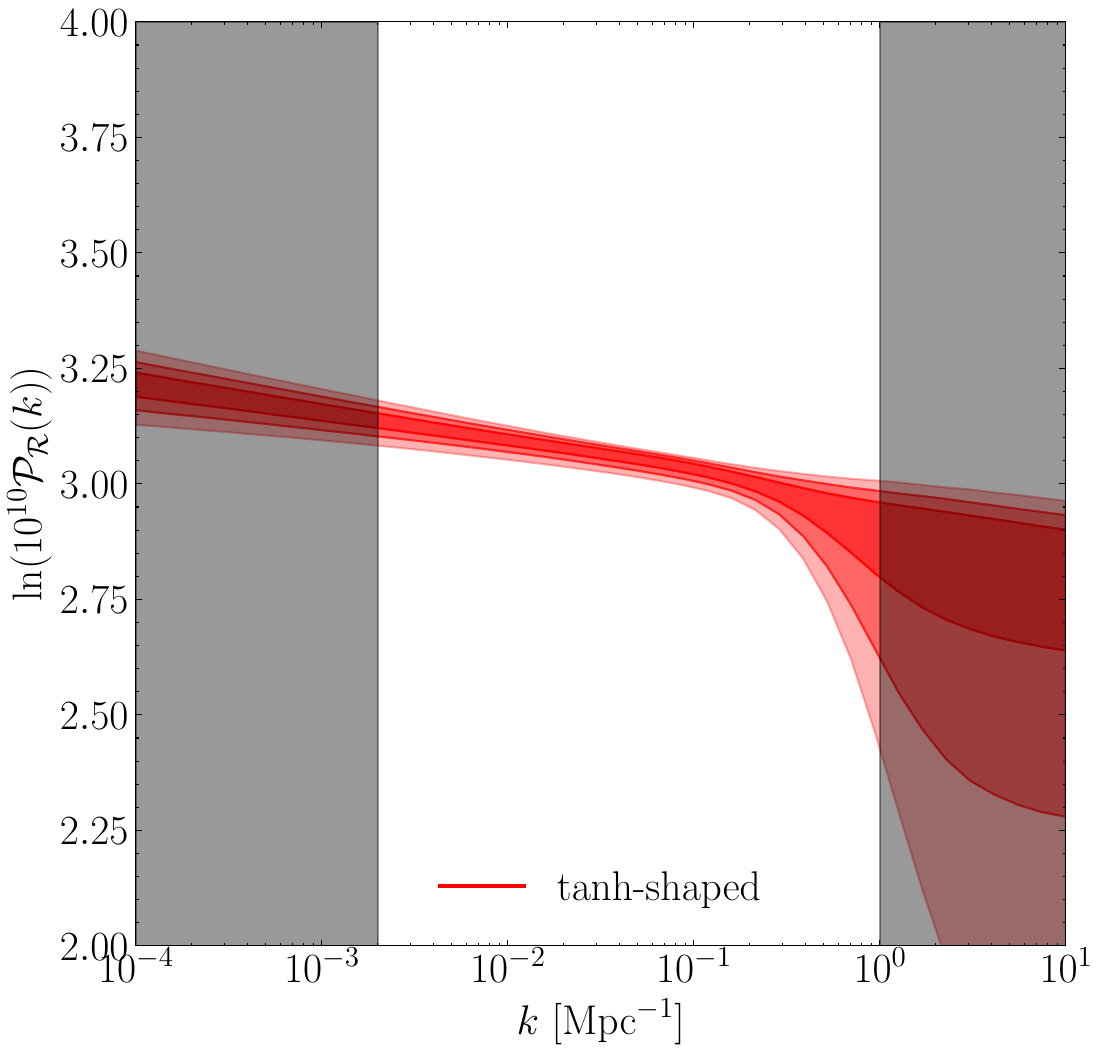}
    \includegraphics[width=\columnwidth]{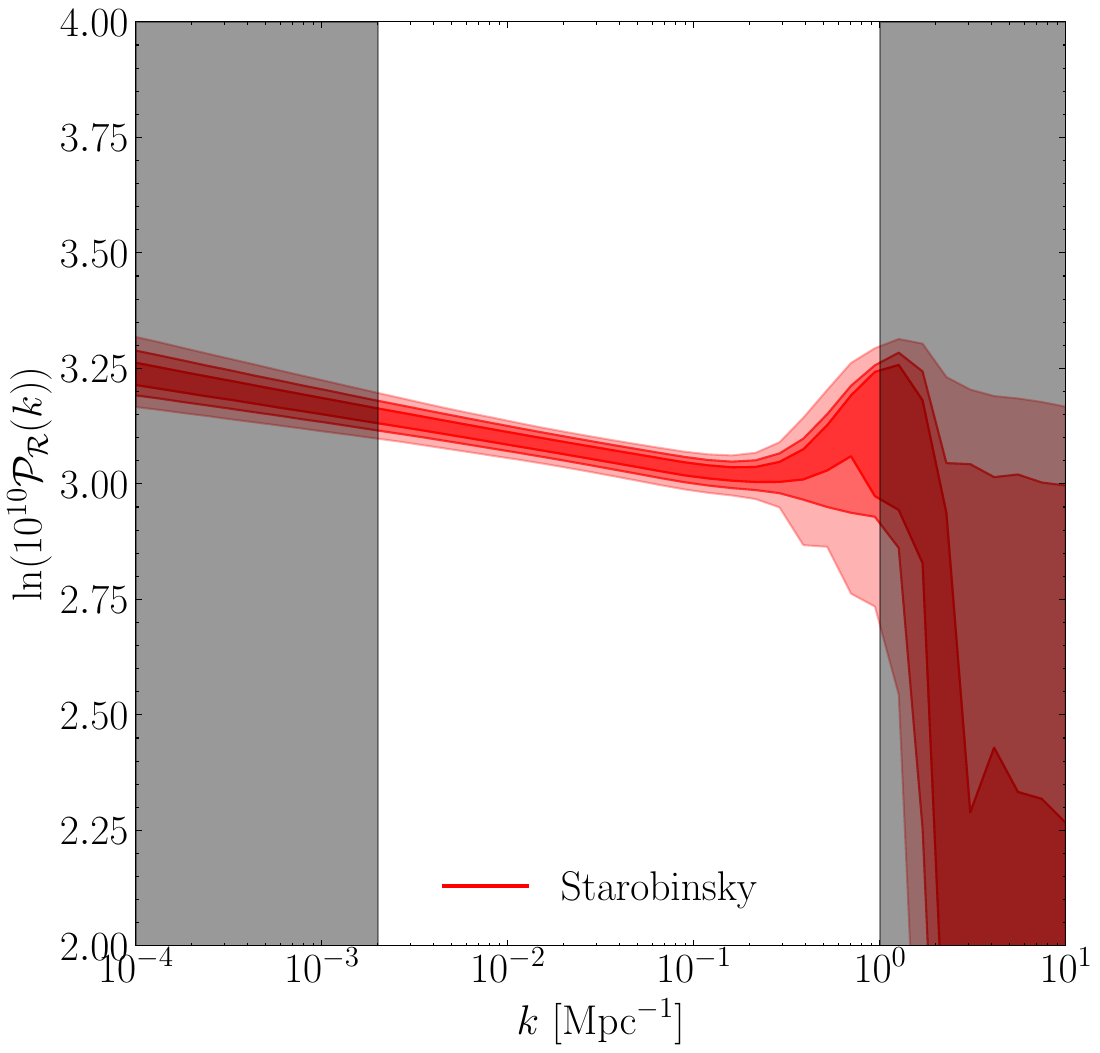}
    \caption{The red-color shaded regions are the 68\%, 95\%, and 99\% credible intervals of 
    the model predictions {of PPS}
    in each $k$ bin, 
    computed from the posterior distribution of the parameter inference
    shown in Figs.~\ref{fig:2D_nrun}, \ref{fig:2D_BPL},~\ref{fig:2D_tanh} and~\ref{fig:2D_SB}). 
    We show the posteriors for the 4 
    {modified PPS models:}
    running spectral index model (upper-left), broken power-law model (upper-right), 
    $\tanh$-shaped model 
    (lower-left), and Starobinsky model (lower-right). The unshaded $x$-axis region denotes the scales probed by the joint analysis: $k \sim [2 \times 10^{-3}, 1]~{\rm Mpc}^{-1}$.
    }
    \label{fig:PPS_templates}
\end{figure*}

In Fig.~\ref{fig:PPS_templates}, 
we show the posterior {distributions of the PPS model predictions computed from the posteriors
of the model parameters in the parameter inference.}
The posterior for each of the {modified PPS models is consistent with the single power-law model over the range of}
wavenumbers used to fit the data, $k \simeq [2\times 10^{-3}, 1]~{\rm Mpc}^{-1}$.

\begin{figure}
    \includegraphics[width=\columnwidth]{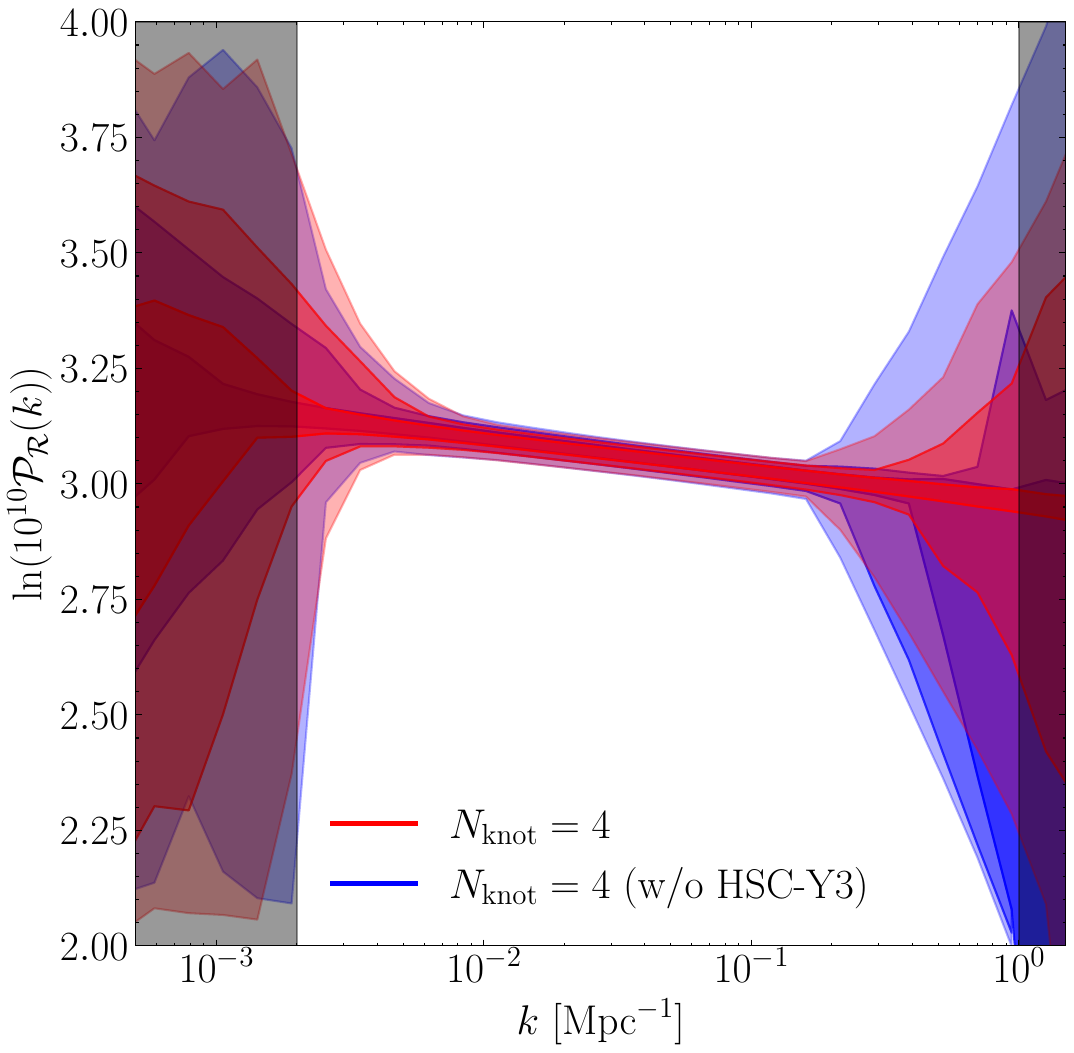}
    \caption{The red-color or blue-color shaded regions are the 68\%, 95\% and 99\% credible intervals of the PPS model predictions in each $k$ bin, 
    computed from the posterior distribution of the parameter inference 
    with or without HSC-Y3 data, respectively.
    {Here, we use the multiple broken power-law PPS model with $N_{\rm knots}=4$, as given 
    in Table~\ref{tab:parameters_free-form}.}
}
    \label{fig:PPS_knots}
\end{figure}

In Fig.~\ref{fig:PPS_knots}, 
we show the posterior of the PPS {from the multiple broken power-law model with $N_{\rm knots}=4$,
corresponding to 2 {breakpoints}
in the $\ln {\cal P}_{\cal R}$-$\ln k$ space. In 
this case, we have 6 parameters to specify the PPS, compared to 4 parameters for the broken power-law model.}
Again {the posterior of PPS is consistent}
with the {single} power-law model over a range of wavenumbers. 
In this figure, {the blue-color posteriors show the results}
without the HSC-Y3 {cosmic shear data.}
Comparing the posteriors with and without the HSC-Y3 data, it is clear that 
{adding the HSC-Y3 data does not affect the PPS constraint over the range}
$k \sim [10^{-2}, 0.1]~{\rm Mpc}^{-1}$, {while it}
improves the constraint on 
smaller scale, $k \gtrsim 0.2~{\rm Mpc}^{-1}$.

{The results in Figs.~\ref{fig:PPS_templates} and \ref{fig:PPS_knots}, which support 
the single power-law model, indicate that the constraint on the PPS is statistically dominated
by the {\it Planck} data, compared to the HSC-Y3 cosmic shear data over the wavenumber range considered.}
Hence, even though the small-scale suppression in the PPS (e.g. Fig.~\ref{fig:PPS_forCS})
{improves consistency between the {\it Planck} {$\Lambda$CDM cosmology} and the HSC-Y3 cosmic shear data,}
the modulation at $k \sim 0.1~{\rm Mpc}^{-1}$ 
{is not favored by the {\it Planck} data in the joint analysis.}

\section{Conclusion}
\label{sec:conclusion}

In this paper, we {performed a joint analysis of}
the HSC-Y3
cosmic shear data, 
the {{\it Planck} CMB data,
the ACT DR6}
CMB lensing data, and the {DESI-Y1}
BAO data, 
to constrain the primordial power spectrum {on small scales while keeping the background cosmology fixed to the flat $\Lambda$CDM model.}

Our findings are summarized as follows:
{
\begin{itemize}
    \item The cosmic shear 2PCFs are sensitive to the modification of the PPS at small scales (Fig.~\ref{fig:Effect_on_2PCFs}) and help to constrain the small-scale PPS when combined with other cosmological probes (Fig.~\ref{fig:PPS_knots}). 
    \item We succeed to constrain the PPS down to $k \sim 1~{\rm Mpc}^{-1}$ and the constraints are consistent with the single power-law PPS model (Figs.~\ref{fig:PPS_templates} and~\ref{fig:PPS_knots}). 
    \item We put constraints on the PPS model parameters (Figs.~\ref{fig:2D_nrun}, \ref{fig:2D_BPL},~\ref{fig:2D_tanh} and~\ref{fig:2D_SB}).
    Those constraints are also consistent with {the single power-law PPS model.}
\end{itemize}
}

{Although our study is {partly} motivated by the $S_8$ tension \citep{wlreana_catalogs_Longley2023,Terasawa24}, we conclude that it cannot be easily resolved by modifying the PPS as long as the background cosmology is fixed within the framework of the adiabatic 
$\Lambda$CDM model. This is because the {\it Planck} data has sufficient statistical power
to
constrain the shape of PPS up to $k\sim 0.2~{\rm Mpc}^{-1}$, which is relevant to the scale
of $S_8${, leaving insufficient freedom of the model to alleviate the $S_8$ tension.}
Therefore, modifying the growth history of matter fluctuations from the $\Lambda$CDM model might provide a better solution, as indicated in Refs.~\cite{2023PhRvL.131k1001N,2024PhRvD.109f3523L}.
We will study this possibility using the HSC-Y3 cosmic shear data in a separate paper {(Terasawa et al., in prep.)}.}

\twocolumngrid

\acknowledgments
We would like to thank Fabian Schmidt for useful discussion. 
This work was supported in part by JSPS KAKENHI Grant Numbers 20H05850,
20H05855,
and 23KJ0747,
and
by World Premier International Research Center Initiative (WPI Initiative), MEXT, Japan.
SS is supported by the JSPS Overseas Research Fellowships. 

\bibliography{refs-short}

\end{document}